\documentstyle[twocolumn,amssymb,aps,psfig,float]{revtex}
\begin{document}
\draft
\twocolumn[\hsize\textwidth\columnwidth\hsize\csname @twocolumnfalse\endcsname

\title{LiVGe$_{2}$O$_{6}$, an anomalous quasi 1D, $S = 1$ system,
as revealed by NMR}

\author{J. L. Gavilano$^{1}$, S. Mushkolaj$^{1}$, H. R. Ott$^{1}$,
P. Millet$^{2}$, and F. Mila$^{3}$}

\address{
$^{1}$ Laboratorium f\"{u}r Festk\"{o}rperphysik,
ETH-H\"{o}nggerberg, CH-8093~Z\"{u}rich, Switzerland \\
$^{2}$ Centre d'Elaboration des Mat$\acute{e}$riaux et d'Etudes
Structurales, 29, rue J. Marvig, 31055 Toulouse Cedex, France \\
$^{3}$ Laboratoire de Physique Quantique, Universit$\acute{e}$ Paul
Sabatier, 31062 Toulouse Cedex, France }


\maketitle

\begin{abstract}
\begin{center} 
\parbox{14cm}
{
We report the results of $^{7}$Li nuclear magnetic resonance (NMR) 
studies of LiVGe$_{2}$O$_{6}$, a quasi one--dimensional spin $S = 1$ 
model system, at low temperatures.  Our data, including NMR spectra 
and the temperature dependence of the spin-lattice relaxation rate 
$T_{1}^{-1}$, indicate a first order phase transition to occur at 
$T_{c}\simeq 23$ K. The NMR response of LiVGe$_{2}$O$_{6}$ below $T_{c}$ 
suggests that the ordered phase is antiferromagnetic and has unusual 
features. Possible reasons for this unexpected behavior are 
discussed.
}
\end{center}
\end{abstract}

\pacs{PACS numbers: 75.30.Et, 75.30.kz, 76.60.-k}
\vskip2pc
]


In recent years there has been considerable progress in the
understanding of the physics of low-dimensional spin systems with an
antiferromagnetic interaction between nearest-neighbor spins.  For a
quasi one-dimensional (1D) Heisenberg chain with integer spins, it is
expected that the ground state is isolated from the excited
states by an energy gap, the Haldane gap
$\Delta_{H}$~\cite{Haldane83}.  This has indeed been observed in many
systems~\cite{renard}.  For the case of monoclinic AgVP$_{2}$S$_{6}$,
where the V$^{3+}$ ions with $S = 1$ form 1D chains along the
$a$-axis, $\Delta_{H}$ is equal to 26 meV~\cite{Mutka91,Takigawa95}.
A continuous spectrum of excitations is expected, however, if
the spins are half-integers.  In this case the system may
lower its energy via a spin-Peierls second order phase transition,
again resulting in the opening of a gap, as observed for CuGeO$_{3}$\cite{Hase}. 
Here, the localized spins of the Cu$^{2+}$
ions form 1D Heisenberg chains.  A spin-Peierls transition is found at
$T_{SP}= 14$ K, the resulting gap is of the order of 2 meV.

Last year a new member, LiVGe$_{2}$O$_{6}$, was added to the growing 
list of 1D magnets by Millet and coworkers~\cite{Millet99}.  From the 
temperature dependence of the magnetic susceptibility $\chi(T)$ it was 
inferred that this $S = 1$ system does not behave as expected.  A 
discontinuous change in $d\chi/dT$ at approximately 22 K was 
interpreted as a phase transition of the spin-Peierls type, and 
quantum chemistry calculations indicated ~\cite{Millet99,Mila99} that 
this unexpected behavior might be due to the presence of a substantial 
biquadratic exchange interaction between the V$^{3+}$ spins forming 
the infinite 1D chains.  From these existing experimental data a more 
detailed analysis of the phase transition and of the nature of the low 
temperature phase was not possible.

In this Letter, we present a detailed $^{7}$Li nuclear magnetic 
resonance (NMR) study invoking NMR spectra and spin-lattice relaxation 
rates.  We also present new data on the magnetic susceptibility of 
LiVGe$_{2}$O$_{6}$.  We confirm that this system ought to be regarded 
as an $S =1$ quasi one-dimensional Heisenberg spin system but the 
character of our NMR data is substantially different from that of 
previously investigated 1D, $S =1$ model systems 
~\cite{Takigawa96,Fujiwara92}.  Our results reveal very unusual low 
temperature properties, including a phase transition at 23 K which is 
not, as previously suggested, a second-order spin--Peierls transition 
to a dimerized phase, but rather a first--order phase transition to a 
magnetically ordered phase, rather unusual for a 1D, $S = 
1$ system.

LiVGe$_{2}$O$_{6}$ crystallizes with a monoclinic structure, space 
group P2$_{1}$/c~\cite{Millet99}.  The structure consists of isolated 
chains of VO$_{6}$ octahedra, joined at the edges.  These chains are 
linked together and kept apart by double chains of distorted GeO$_{4}$ 
tetrahedra, keeping the interchain coupling small.  The Li atoms are 
six-fold coordinated and are located inside distorted oxygen 
octahedra as in the case of LiVSi$_{2}$O$_{6}$~\cite{Satto97}. Our 
powder sample of 0.13 g was prepared as described by Millet et 
al~\cite{Millet99}.

\begin{figure}[ht]
\vspace{2ex plus 1ex minus 0.5ex}
\centerline{\psfig{figure=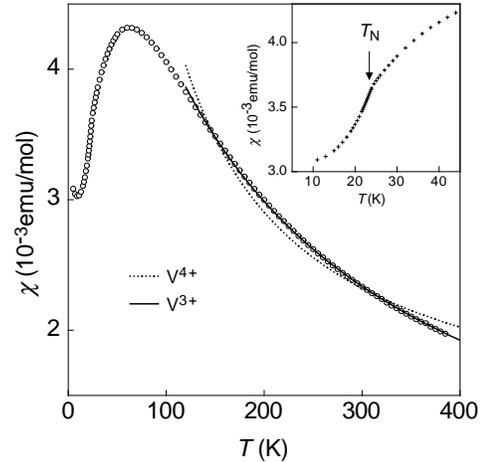,width=7.0cm,angle=0}}
\caption{
Magnetic susceptibility $\chi$ as a function of temperature for
LiVGe$_{2}$O$_{6}$.  The solid line represents the best fit to the
data assuming a Curie-Weiss law for V$^{3+}$ and the dotted line the
corresponding fit for V$^{4+}$.  The inset  displays $\chi(T)$ in the
region near the phase transition.  The arrow indicates $T_{c}$.
}
\protect
\label{Figure1}
\end{figure}

\begin{figure}[ht]
\vspace{2ex plus 1ex minus 0.5ex}
\centerline{\psfig{figure=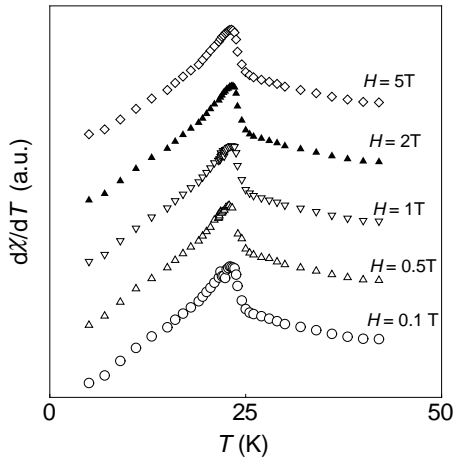,width=7.0cm,angle=0}}
\caption{
$d\chi/dT$ as a function of the temperature for different
applied fields.  The clear maximum in $d\chi/dT(T)$ signals the
transition temperature $T_{c}$.
}
\protect
\label{Figure2}
\end{figure}

An important issue concerning the basic physics of LiVGe$_{2}$O$_{6}$ 
is to verify that the V ions are in a trivalent oxydation state and 
hence $S = 1$.  This claim is supported by our results for the 
electrical resistivity $\rho$ which, measured on a sample prepared 
from pressed powder of LiVGe$_{2}$O$_{6}$, was found to be larger than 
$2 \times 10^{7}$ $\Omega$-cm at room temperature, indicating that 
LiVGe$_{2}$O$_{6}$ is indeed an insulator.  Taking into account the 
known oxydation states of O$^{2-}$, Li$^{1+}$, and Ge$^{4+}$, the 
valence of V must be $+3$.  Further support for this simple, yet 
important clarification is provided by the results of the magnetic 
susceptibility $\chi(T)$ (Fig.  1).  Our results are similar to those 
of Ref.  \cite{Millet99}, albeit with a smaller paramagnetic 
contribution at low temperatures.  The high temperature part of 
$\chi(T)$ can reasonably be fitted only by assuming a trivalent 
configuration of V. The phase transition, suggested by a kink in 
$\chi(T)$ at $T_{c}= 23$ K (see inset to Fig.  1), is much more 
evident by plotting $d\chi/dT$ versus $T$, as demonstrated 
in Fig.~\ref{Figure2}.  The well defined maximum is only weakly, if at 
all, shifted by an external magnetic field.  The upturn of $\chi(T)$ 
at the lowest temperatures is associated with a small amount of 
paramagnetic impurities (a few parts per mil of $S = 3/2$) and will 
not be considered further.

\begin{figure}[ht]
\vspace{2ex plus 1ex minus 0.5ex}
\centerline{\psfig{figure=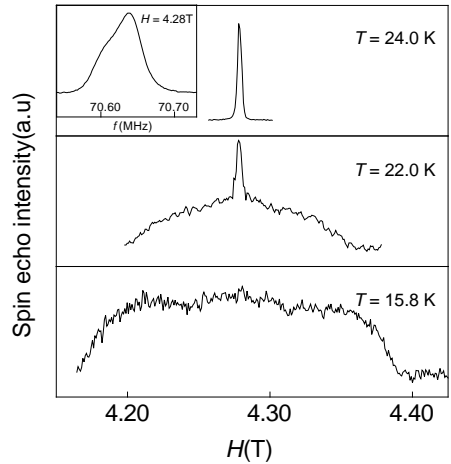,width=7.0cm,angle=0}}
\caption{ $^{7}$Li NMR spectra of LiVGe$_{2}$O$_{6}$ measured at a 
fixed frequency of 70.64 MHz for three different temperatures.  Below 
23 K the $^{7}$Li-NMR spectrum becomes very broad.  The inset to the 
top figure shows the $^{7}$Li NMR spectrum measured at a fixed applied 
field of 4.28 T and 39 K.
}
\protect
\label{Figure3}
\end{figure}

In Fig.~\ref{Figure3} we show three examples of the $^{7}$Li NMR 
(nuclear spin $I = 3/2$) spectra measured at a fixed frequency of 70.64 MHz 
and at temperatures of 15.8, 22.0 and 24 K. For these measurements a 
two-pulse spin-echo sequence was employed and the data represents the 
integrated spin-echo intensity.  An example of the NMR spectrum for 
randomly oriented powder of LiVGe$_{1}$O$_{6}$ at 39 K is shown in the 
inset of Fig.  3.  This data was acquired at a fixed applied field of 
4.28 T. This spectrum reveals the absence of quadrupolar wings, and 
its shape indicates an anisotropic shift.  The maximum intensity and 
the prominent shoulder correspond to grains where the Li sites have 
their quadrupolar axis orthogonal and parallel to the applied field, 
respectively.  Below 30 K the shape of the signal develops into a well defined 
and symmetrical line with a width (HWHM) of approximately 35 
Gauss (see the NMR spectrum at 24 K), which we take as evidence for 
the alignment of the grains of our powdered sample in the presence of applied 
magnetic fields of the order of a few Tesla at low temperatures.  
Another scenario where the observed changes in the NMR spectrum simply 
reflect changes in the susceptibility and/or hyperfine field 
coupling, yielding an isotropic line shift below 30 K, cannot really 
be ruled out.

From the NMR line shift $K$ data (not shown here) we estimate the 
hyperfine field at the Li nuclei to be of the order of 0.65 kG per 
$\mu_{B}$ of V magnetic moment.  This value seems consistent with a 
direct dipolar coupling between the V magnetic moments and the Li 
nuclei and we conclude that the $^{7}$Li NMR response is dominated by 
the magnetism of the V ions.  The temperature evolution of the 
NMR spectrum above 23 K shows no indication of a gap in the spectrum 
of excitations of the spin system.

At temperatures below $T_{c}$ the NMR spectrum changes dramatically
and its width increases very rapidly with decreasing $T$.  At
temperatures not far below $T_{c}$ (see the spectrum for 22 K in
Fig.  3) we note the coexistence of two phases.  The narrow line,
related with the high-temperature paramagnetic phase, appears on top of
a broad background representing the low-temperature phase.  At even
lower temperatures only the signal due to the low temperature phase is
measured.  This behavior is compatible with a first order phase
transition at $T_{c}$ but not with a second order phase
transition such as a standard spin-Peierls transition or a simple
magnetic ordering phenomenon. In particular, one also may argue that 
the coexistence of the two phases is due to a spread of transition 
temperatures within the sample material\cite{MacLauglin94}. In the 
latter case it would seem quite unlikely to observe the distinct peak in 
the temperature variation of $T_{1}^{-1}$, displayed in Fig. 4.

On the basis of the presently available experimental data, nothing 
definitive can be said concerning the nature of the low temperature 
phase. The broad NMR spectrum may either represent the powder 
spectrum of a distorted antiferromagnetically ordered phase or, in case 
that the powder grains are indeed aligned at low temperature, a 
modulated magnetic structure is also conceivable. Further experiments 
on single crystals will have to answer this question.

In order to probe the low-energy spin excitations in 
LiVGe$_{2}$O$_{6}$ at low temperatures, we measured the spin-lattice 
relaxation rate $T_{1}^{-1}$ by monitoring the recovery of the 
$^{7}$Li nuclear magnetization after the application of a long comb of 
$rf$ pulses above and below $T_{c}$.  Both above and below $T_{c}$ a 
single exponential recovery was observed, the first as expected and 
 the latter in spite of the 
fact that the NMR spectrum is very broad and hence cannot fully be 
irradiated.  Various changes of irradiation conditions yielded, within 
the usual error limits, identical results for $T_{1}^{-1}$, however.

\begin{figure}[ht]
\vspace{2ex plus 1ex minus 0.5ex}
\centerline{\psfig{figure=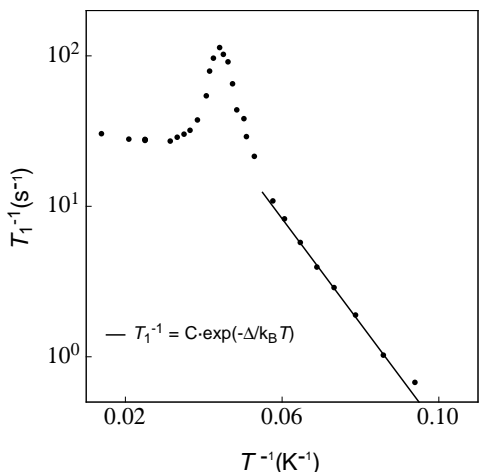,width=7.0cm,angle=0}}
\caption{
Temperature dependence of the $^{7}$Li spin-lattice relaxation rate in
LiVGe$_{2}$O$_{6}$. Above 28 K $T_{1}^{-1}$ is only weakly
$T-$dependent, but at low temperatures $T_{1}^{-1}$ decreases
exponentially with decreasing temperatures.
}
\protect
\label{Figure4}
\end{figure}

In Fig.~\ref{Figure4} we display the temperature dependence of the 
spin-lattice relaxation rate.  A prominent peak in $T_{1}^{-1}(T)$ 
reflects the phase transition at $T_{c}$.  Above $T_{c}$, $T_{1}^{-1}$ 
varies only weakly with temperature, a further evidence for the 
absence of a significant energy gap in the spin excitation spectrum.  
For an insulator the relaxation rate is very high and is most likely 
caused by fluctuations of the localized V-ion moments.  Below $T_{c}$, 
$T_{1}^{-1}$ decreases very rapidly, signalling the opening of a gap 
in the spectrum of spin excitations.  From the relaxation data between 
17 and 10 K, we calculate a gap $\Delta/k_{B}$ = 83 K, obviously a 
feature of the low temperature phase.

The magnitude of this gap is very surprising because it is 
substantially larger than the estimated exchange interaction $J$ above 
$T_{c}$.  For standard $S = 1$ chain systems, the maximum of the 
susceptibility at 60 K would imply that, for the high temperature 
phase, $J/k_{B} \approx 45$ K. This suggests that some kind of 
modification, presumably involving the crystal structure, takes place 
at the transition, changing the local quantum chemistry and enhancing 
the exchange interaction.  In case of a standard transition to 
magnetic order with $J/k_{B} \approx$ 45 K, and recalling that the 
anisotropy $D/k_{B}$ is smaller than 20 K\cite{Millet99}, the 
spin-orbit gap would be of the order of $(2DJ/k_{B}^{2})^{1/2}$, i.e., 
at most 40 K.

The unexpected properties of LiVGe$_{2}$O$_{6}$, not typical for quasi 
1D, $S = 1$ systems, are very likely related to the three $t_{2g}$ 
orbitals accommodating the two $3d$ electrons of V$^{3+}$, as for 
V$_2$O$_3$\cite{Rice}.  The orbital $d_{xz}$, which is separated from 
the other two by a second order crystal-field splitting $\Delta_{CF}$ 
due to the trigonal distortion\cite{Millet99}, may modify the usual 
description in terms of a Heisenberg model in two ways.

If $\Delta_{CF}$ significantly exceeds the hopping integrals between 
neighbouring sites, the low-energy physics can still be described by a 
pure spin Hamiltonian.  However, the third orbital gives rise to a 
ferromagnetic exchange channel.  The Heisenberg coupling may thus 
considerably be reduced and other exchange couplings, such as a 
biquadratic interaction $J''$ between nearest neighbours or a bilinear 
exchange $J_{2}$ between next-nearest neighbours may become 
significant.  In that case, the high temperature magnetic properties 
of the system would be described by the Hamiltonian
\begin{equation}
         H =  \sum_{i}^{} \left [ J' {\bf S}_{i}\cdot{\bf
         S}_{i+1} + J'' ({\bf S}_{i}\cdot{\bf
S}_{i+1})^{2} + J_2 {\bf S}_{i}\cdot{\bf S}_{i+2} \right ] .
\label{equation1}
\end{equation}

While the properties of the Hamiltonian of Eq.~(\ref{equation1}) with 
a frustrating next-nearest neighbour coupling have not been 
investigated yet, it is likely that the frustration due to $J_2$ will 
lead to incommensurate fluctuations, as in the case 
$J''=0$\cite{Kolezhuk97}, or the biquadratic interaction $J''$ will 
tend to close the Haldane gap, as in the case of 
$J_2=0$\cite{Affleck87,Schollwock96}.  With $J''/J'\simeq 1$ and 
$J_2/J'>0.35$, we conjecture that this Hamiltonian leads to a very 
small gap and to incommensurate fluctuations.  This might provoke a 
spin-Peierls type instability to an incommensurate phase and explain part 
of our results.  Whether the transition can be of first order remains 
to be seen.

If $\Delta_{CF}$ adopts similar values as the hopping integrals, it 
will no longer be possible to describe the low energy properties of 
the system with a pure spin Hamiltonian, and the orbital degrees of 
freedom have to be included explicitly.  The canonical example is 
V$_2$O$_3$, and the properties reported here are reminiscent of the 
phase transition observed in insulating V$_{2-x}$Cr$_{x}$O$_{3}$ 
between the high-temperature paramagnetic and the low-temperature 
antiferromagnetic phase\cite{Rice}.  In particular, this transition, 
due to an orbital ordering of the $t_{2g}$ orbitals, is strongly 
first-order and there is a dramatic change of exchange integrals at 
the transition.  If an orbital ordering takes place in 
LiVGe$_{2}$O$_{6}$, it will quite likely also provoke a first order 
transition.  The nature of the low temperature phase will depend on 
the effective spin Hamiltonian, hence on the orbital ordering.  The 
possibilities range from a dimerized ground state involving a simple 
antiferro-orbital ordering with alternating exchange couplings, to 
more exotic phases including incommensurate phases if the orbital 
ordering is helical.  In that case, the large gap would just be the 
spin-orbit gap, the exchange integrals being much larger than in the 
high temperature phase because of the orbital ordering.  Further 
investigations testing the orbital occupancy on the V$^{3+}$ ions are 
clearly needed to check this possibility.  Unfortunately the shape of 
the low temperature NMR spectra does not allow for firm conclusions 
with regard to orbital ordering.

In summary, our results for LiVGe$_{2}$O$_{6}$ strongly suggest 
that this material is a very unusual 1D, $S = 1$, Heisenberg system.  
In particular the Haldane phase is either absent or strongly 
suppressed, and a first order phase transition into a magnetically 
ordered phase occurs at 23 K. On the theoretical front we have 
argued that this behavior is probably related to the second order 
splitting of the $t_{2g}$ orbitals, which could either induce 
significant biquadratic and next-nearest neighbor exchange 
interactions along the chain, or provide an orbital degree of freedom 
which is involved in the ordering.  Further investigations of the 
low-temperature phase and of the theoretical models are obviously needed 
to answer the questions that have been raised by our observations.

F. M. acknowledges useful discussions with D. Poilblanc, E. Sorensen, 
F.-C. Zhang and Y. Fagot-Revurat.  This work was financially supported 
by the Schweizerische Nationalfonds zur F\"{o}rderung der 
Wissenschaftlichen Forschung.


\end{document}